\documentclass{aa}

\usepackage{epsfig,astron}

\begin{document}

\thesaurus{ 08 (09.03.1 ; 09 04 1 ; 09.19.1 )}

\title{Mapping of the extinction in Giant Molecular Clouds using optical star
counts}

\author{L. Cambr\'esy} 
          
\institute{Observatoire de Paris, D\'epartement de Recherche Spatiale, F-92195 Meudon Cedex, France} 
\offprints{Laurent Cambr\'esy,\\  Laurent.Cambresy@obspm.fr}
            
\date{}

\titlerunning{Mapping of the extinction using star counts}
\maketitle

\begin{abstract}
This paper presents large scale extinction maps of most nearby Giant Molecular
Clouds of the Galaxy (Lupus, $\rho$ Ophiuchus, Scorpius, Coalsack, Taurus,
Chamaeleon, Musca, Corona Australis, Serpens, IC 5146, Vela, Orion, Monoceros
R1 and R2, Rosette, Carina) derived from a star count method using an adaptive
grid and a wavelet decomposition applied to the optical data provided by the 
USNO-Precision Measuring Machine. 
The distribution of the extinction in the clouds leads to estimate their  total individual
masses  $M$ and their maximum of extinction. I show that the relation between
the mass contained within  an iso--extinction contour and the extinction is 
similar from cloud to cloud and allows the extrapolation of the maximum of 
extinction in the range 5.7 to 25.5 magnitudes. I found that about half of 
the mass is contained in regions where the visual extinction is smaller than
1 magnitude. 
The star count method used on large scale ($\sim 250$ square degrees) is a
powerful and relatively straightforward method to estimate the mass of
molecular complexes.
A systematic study of the all sky would lead to discover new clouds as I
did in the Lupus complex for which I found a sixth cloud of about
$10^4 \, M_{\sun}$.

\end{abstract}

\section{Introduction}
Various methods have been recently developed to estimate the mass of matter
contained in giant molecular clouds (GMC) using millimetric and far infrared
observations \cite{BBDT98,MHY+98}. 
Nevertheless, the mapping of the optical/near--infrared extinction, based on
star counts still remain the most straightforward way to estimate the mass in
form of dust grains. These maps can be usefully compared to longer wavelength
emission maps in order to derive the essential physical parameters of the
interstellar medium such as the gas to dust mass ratio, the clumpiness of the
medium or the optical and morphological properties of the dust grains.
The star count method was first proposed by Wolf \cite*{Wol23} and has been
applied to Schmidt plates during several decades. It consists to count the
number of stars by interval of magnitudes (i.e. between $m-1/2$ and $m+1/2$)
in each cell of a regular rectangular grid in an obscured area and to compare 
the result with the counts obtained in a supposedly unextinguished region. In
order to improve the spatial resolution, Bok \cite*{Bok56} proposed to make
count up to the completeness limiting magnitude ($m \le m_{lim}$). Since the 
number of stars counted is much larger when compared to counts performed in an
interval of 1 magnitude, the step of the grid can be reduced. Quite recently
extinction maps of several southern clouds have been drawn by Gregorio Hetem
et al. \cite*{GSL88} using this second method. Even more recently, Andreazza and
Vilas-Boas \cite*{AV96} obtained extinction map of the Corona Australis and 
Lupus clouds. Counts were done {\em visually} using a $\times 30$ magnification
microscope. With the digitised Schmidt plates the star counts method can be
worked out much more easily across much larger fields. The star count methods
have also tremendously evolved thanks to the processing of digital data with
high capacity computers. Cambr\'esy et al. \cite*{CEC+97}, for instance, have
developed a counting method, for the DENIS data in the Chamaeleon I cloud,
 that takes advantage of this new environment. 
The aim of this paper is to apply this method to the optical plates digitised
with the USNO-PMM in a sample of Giant Molecular Clouds to derive their 
extinction map (Vela, Fig. \ref{vela};
Carina, Fig. \ref{carina};
Musca, Fig. \ref{musca};
Coalsack, Fig. \ref{coalsack};
Chamaeleon, Fig. \ref{cha};
Corona Australis, Fig. \ref{crA};
IC 5146, Fig. \ref{ic5146};
Lupus, Fig. \ref{lupus};
$\rho$ Ophiuchus, Fig. \ref{rho_oph};
Orion, Fig. \ref{orion};
Taurus, Fig. \ref{taurus};
Serpens, Fig. \ref{serpens}).
The method is shortly described in Sect. \ref{method}. In Sect. \ref{deduced},
deduced parameters from the extinction map are presented and 
Sect. \ref{remarks} deals with individual clouds.

\section{Star counts}
\label{method}
\subsection{Method}
The star counts method is based on the comparison of local stellar densities.
A drawback of the classical method is that it requests a grid step. If the step
is too small, this may lead to empty cells in highly extinguished regions, and
if it is too large, it results in a low spatial resolution. My new approach
consists in fixing the number of counted stars per cell rather than the step of
the grid. Since uncertainties in star counts follow a poissonian distribution,
they are independent of the local extinction with an adaptive grid where the
number of stars in each cell remains constant. 
Practically, I used a fixed number of 20 stars per cell and a filtering method
involving a wavelet decomposition that filters the noise. This method has been
described in more details by Cambr\'esy \cite*{Cam98a}.

\begin{figure}[htb]
\caption{Extinction map of Vela from $B$ counts (J2000 coordinates)}
\label{vela}
\end{figure}

\begin{figure}[htb]
\caption{Extinction map of Carina from $R$ counts (J2000 coordinates)}
\label{carina}
\end{figure}

\begin{figure}[htb]
\caption{Extinction map of Musca from $R$ counts (J2000 coordinates)}
\label{musca}
\end{figure}

I have applied the method to 24 GMCs. Counts and filtering are fully
automatic.
Because of the wide field ($\sim 250$ square degrees for Orion), the counts
must be corrected for the variation of the background stellar density with 
galactic latitude. Extinction and stellar density are related by:
\begin{equation}
A_\lambda = \frac{1}{a} \log \left( \frac{D_{ref}(b)}{D} \right)
\label{extinction}
\end{equation}

\noindent
where $A_\lambda$ is the extinction at the wavelength $\lambda$, $D$ is the
background stellar density, $D_{ref}$ the density in the reference field
(depending on the galactic latitude $b$), and $a$ is defined by :
\begin{equation}
a = \frac{\log (D_{ref}) - cst.}{m_\lambda}
\label{lum_funct}
\end{equation}

\noindent
where $m_\lambda$ is the magnitude at the wavelength $\lambda$.

\noindent
Assuming an exponential law for the stellar density, 
$D_{ref}(b)=D_0 \, {\rm e}^{-\alpha|b|}$, a linear correction with the
galactic
latitude $b$ must be applied to the extinction value given by
Eq.~(\ref{extinction}).
The correction consists, therefore, in subtracting 
$\log [D_{ref}(b)] = \log (D_0) - \alpha |b| \, \log {\rm (e)}$
to the extinction value $A_\lambda(b)$. This operation corrects the slope of
$A_\lambda(b)$ which becomes close to zero, and set the zero point of
extinction. All maps are converted into visual magnitudes assuming an 
extinction law of Cardelli et al. \cite*{CGM89} for which
$\frac{A_B}{A_V} = 1.337$ and $\frac{A_R}{A_V}=0.751$.

Here, the USNO-PMM catalogue \cite{Mon96} is used to derive the extinction
map. It results from the digitisation of POSS (down to $-35\degr$ in
declination) and ESO plates ($\delta \le -35\degr$) in blue and red.
Internal photometry estimators are believed to be accurate to about 0.15
magnitude but systematic errors can reach 0.25 magnitude in the North and
0.5 magnitude in the South. Astrometric error is typically of the order of
0.25 arcsecond. This accuracy is an important parameter in order to count only
once those stars which are detected twice because they are located in the 
overlap of two adjacent plates.

All the extinction maps presented here have been drawn in greyscale with
iso--extinction contours overlaid. On the right side of each map, a scale
indicates the correspondence between colours and visual extinction, and the
value of the contours. Stars brighter than the $4^{th}$ visual magnitude 
\cite{HJ91} are marked with a filled circle.

\begin{figure*}[htb]
\caption{Extinction map of Coalsack from $R$ counts (J2000 coordinates)}
\label{coalsack}
\end{figure*}

\begin{figure*}[htb]
\caption{Extinction map of the Chamaeleon complex from $B$ counts (J2000 coordinates)}
\label{cha}
\end{figure*}

\begin{figure*}[htb]
\caption{Extinction map of Corona Australis from $B$ counts (J2000 coordinates)}
\label{crA}
\end{figure*}

\begin{figure*}[htb]
\caption{Extinction map of IC5146 from $R$ counts (J2000 coordinates)}
\label{ic5146}
\end{figure*}

\begin{figure*}[htb]
\caption{Extinction map of the Lupus complex from $B$ counts (J2000 coordinates)}
\label{lupus}
\end{figure*}

\begin{figure*}[htb]
\caption{Extinction map of $\rho$ Ophiuchus and Scorpius clouds from $R$
counts (J2000 coordinates)}
\label{rho_oph}
\end{figure*}

\subsection{Artifacts}
Bright stars can produce {\em artifacts} in the extinction maps. A very bright
star, actually, produces a large disc on the plates that prevents the
detection of the fainter star (for example, $\alpha$ Crux in the Coalsack or
{\em Antares} in $\rho$ Ophiuchus, in Fig. \ref{coalsack} and \ref{rho_oph},
respectively).  The magnitude of
{\em Antares} is $m_V=0.96$, and it shows up in the extinction map as a disc 
of 35\arcmin\ diameter which mimics an extinction of 8 magnitudes. Moreover
{\em Antares} is accompanied of reflection nebulae that prevent source
extraction. In Fig. \ref{orion} the well known Orion constellation is drawn 
over the map and the brighter stars appear. $\epsilon \, Ori$ ({\em Alnilam})
the central star of the constellation, free of any reflection nebula, is 
represented by a disc of $\sim 14\arcmin$ for a magnitude of $m_V = 1.7$. 
Fortunately, these artifacts can be easily identified when the bright star is 
isolated. When stars are in the line of sight of the obscured area, the
circularity of a small extinguished zone is just an indication, but the only
straightforward way to rule out a doubt is to make a direct visual inspection
of the Schmidt plate. 

Reflection nebulae are a more difficult problem to identified since they are
not always circular. Each time it was possible, I chose the $R$ plate 
because the reflection is much lower in $R$ than in $B$. $B$ plates were
preferred when $R$ plates showed obvious important defaults (e.g. edge of the
plate). 

\subsection{Uncertainties}
Extinction estimations suffer from intrinsic and systematic errors.
Intrinsic uncertainties result essentially from the star counts itself. The
obtained distribution follows a Poisson law for which the parameter is
precisely the number of stars counted in each cell, i.e. 20.
Equation (\ref{extinction}) shows that two multiplicative factors, depending
on which colour the star counts is done, are needed to convert the stellar 
density into visual extinction. These factors are $a$ (the slope of the 
luminosity function (\ref{lum_funct})), and the conversion factor 
$A_\lambda / A_V$. For $B$ and $R$ band, the derived extinction accuracies are
$^{+0.29}_{-0.23}$ magnitudes and $^{+0.5}_{-0.4}$ magnitudes, respectively.

Also, for highly obscured region, densities are estimated from counts on large
surfaces, typically larger than $\sim 10'$. It is obvious that, in this case,
the true peak of extinction is underestimated since we have only an average
value. The resulting effect on the extinction map is similar to the 
{\em saturation} produced by bright stars on Schmidt plates. This effect 
cannot be easily estimated and is highly dependant on the cloud (about
$\sim 80$ magnitudes in $\rho$ Ophiuchus, see Sect. \ref{max_ext}).

Moreover, systematic errors due to the determination of the zero point of
extinction are also present. Extinction mappings use larger areas than the 
cloud itself in order to estimate correctly the zero point. This systematic
uncertainty can be neglected in most cases.

\section{The extinction maps and the derived parameters}
\label{deduced}
\begin{table}
\caption{Cloud properties. Distances are taken from literature,
masses (expressed in solar masses) are defined by the regression line
$\log M(A_V)~=~\log (M_{\rm Tot})~+~a~\times~A_V$, the maxima of extinction, 
$A^m_{V}$, is measured from star counts, and 
$A^e_{V}$, is extrapolated from the previous equation assuming a
fractal structure and the last column is the value of the slope $a$}
\label{tab}
\begin{tabular}{p{1.8cm}rlrrrr}
\hline
Cloud Name & d (pc) & $A^m_{V}$&$A^e_{V}$ & $M_{Tot}$ & Slope \\
\hline
Lupus I          & 100$^{(1)\ \ \ }$ & 5.3 & 7.1 &       $10^4$ &-0.56 \\
Lupus II         & 100$^{(1)\ \ \ }$ & 3.8 & 5.7 &          80  &-0.33 \\
Lupus III        & 100$^{(1)\ \ \ }$ & 4.9 & 7.6 &        1150  &-0.40 \\
Lupus IV         & 100$^{(1)\ \ \ }$ & 5.3 & 7.0 &         630  &-0.40 \\
Lupus V          & 100$^{(1)\ \ \ }$ & 5.2 & 10.6&        2500  &-0.32 \\
Lupus VI         & 100$\ \ \ \ \ \ $ & 4.8 & 7.0 &       $10^4$ &-0.57 \\
$\rho$ Ophiuchus & 120$^{(1)\ \ \ }$ & 9.4 & 25.5&        6600  &-0.15 \\
Scorpius         & 120$\ \ \ \ \ \ $ & 6.4 & 7.0 &        6000  &-0.54 \\
Taurus           & 140$^{(2)\ \ \ }$ & 7.5 & 15.7&$1.1 \, 10^4$ &-0.26 \\
Coalsack         & 150$^{(1)(3)}$    & 6.6 & 6.3 &$1.4 \, 10^4$ &-0.63 \\
Musca            & 150$^{(1)\ \ \ }$ & 5.7 & 10.1&         550  &-0.27 \\
Chamaeleon~III   & 150$^{(4)\ \ \ }$ & 3.7 & 7.8 &        1300  &-0.40 \\
Chamaeleon~I     & 160$^{(4)\ \ \ }$ & 5.2 & 12.9&        1800  &-0.25 \\
Corona\-Australis& 170$^{(1)\ \ \ }$ & 5.4 & 10.7&        1600  &-0.30 \\
Chamaeleon~II    & 178$^{(4)\ \ \ }$ & 4.9 & 12.3&         800  &-0.22 \\
Serpens          & 259$^{(5)\ \ \ }$ & 10.1& ??? &         ???  &\ \ ??? \\
IC 5146          & 400$^{(6)\ \ \ }$ & 6.5 & 11.9&        2900  &-0.29 \\
Vela             & 500$^{(7)\ \ \ }$ & 4.0 & ??? &         ???  &\ \ ??? \\
Orion            & 500$^{(8)\ \ \ }$ & 7.5 & 20.3& $3 \, 10^5$  &-0.27 \\
Crossbones       & 830$^{(8)\ \ \ }$ & 4.4 & 10.4&$7.3 \, 10^4$ &-0.47 \\
Monoceros~R2     & 830$^{(8)\ \ \ }$ & 4.1 & 10.5&$1.2 \, 10^5$ &-0.48 \\
Monoceros~R1     &1600$^{(9)\ \ \ }$ & 5.4 & 20.8&$2.7 \, 10^5$ &-0.26 \\
Rosette          &1600$^{(9)\ \ \ }$ & 8.4 & 20.4&  $5 \, 10^5$ &-0.28 \\
Carina           &2500$^{(10)\ \ }$ & 8.3 & 82?&$5.5 \, 10^5$?&-0.07?\\
\hline
\end{tabular}

\vspace{1ex}
$(1)$~:~Knude~and~Hog \cite*{KH98},
$(2)$~:~Kenyon~at~al. \cite*{KDH94},
$(3)$~:~Franco \cite*{Fra95},
$(4)$~:~Whittet~et~al. \cite*{WPF+97},
$(5)$~:~Strai\v zys~et~al. \cite*{SCB96},
$(6)$~:~Lada~et~al. \cite*{LLCB94},
$(7)$~:~Duncan~et~al. \cite*{DSHJ96},
$(8)$~:~Maddalena~et~al. \cite*{MMMT86},
$(9)$~:~Turner \cite*{Tur76},
$(10)$~:~Feinstein \cite*{Fei95}
\end{table}

\subsection{Fractal distribution of matter in molecular clouds}
Fractals in molecular clouds characterize a geometrical property which is the
dilatation invariance (i.e. self--similar fractals) of their structure.
A fractal dimension in cloud has been first found in Earth's atmospheric clouds
by comparing the perimeter of a cloud with the area of rain. Then, using
{\em Viking} images, a fractal structure for Martian clouds has also been 
found. Bazell and D\'esert \cite*{BD88} obtained similar results for the 
interstellar cirrus discovered by IRAS.
Hetem and L\'epine \cite*{HL93} used this geometrical approach to generate
clouds with some statistical properties observed in real clouds. They showed 
that classical models of spherical clouds can be improved by a fractal
modelisation which depends only on one or two free parameters. The mass
spectrum of interstellar clouds can also be understood assuming a fractal
structure \cite{EF96}. Larson \cite*{Lar95} went further, showing that the 
Taurus cloud also presents a fractal structure in the distribution of its 
young stellar objects.

\begin{figure}[htb]
	\epsfig{figure=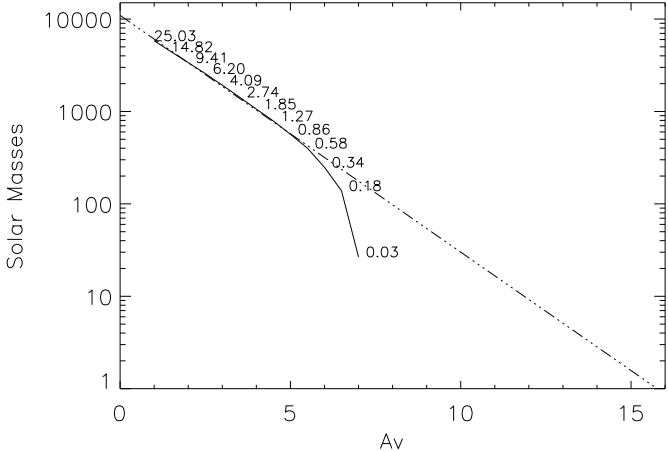,width=8.8cm}
\caption{Mass contained inside the iso--extinction contours versus the extinction (solid line), and regression line for the linear part. Annotations indicate the area in square degrees contained by the iso--extinction contours $A_V$.}
\label{mass_spec}
\end{figure}

Blitz and Williams \cite*{BW97} claim, however, that clouds are no longer
fractal since they found a characteristic size scale in the Taurus cloud.
They showed that the Taurus cloud is not fractal for a size scale of
0.25-0.5 pc which may correspond to a transition from a turbulent outer
envelope to an inner coherent core.
This is not inconsistent with a fractal representation of the cloud for size
scales greater than 0.5 pc. Fractal in physics are defined over a number of 
decades and have {\it always} a lower limit.

Fractal structures in clouds can be characterized by a linear relation between
the radius of a circle and the mass that it encompasses in a log-log diagram.
Several definitions of fractal exist, and this definition can be written
$M \propto L^D$, where $L$ is the radius of the circle and $D$ the fractal
dimension of the cloud.
The mass measured is, in fact, contained in a cylinder of base radius $L$ and
of undefined height $H$ (because the cloud depth is not constant over the
surface of the base of the cylinder).
In our case, we are interested in the relation between the mass and the
extinction. Since the extinction is related to the size $H$, which represents
the depth of the cloud as defined above, we seek for a relation between
mass and $A_V$ (or $H$), $L$ being now undefined.
The logarithm of the mass is found to vary linearly with the extinction over
a range of extinction magnitudes (Fig. \ref{mass_spec}). We have:
\begin{equation}
\log M = \log M_{Tot} + slope \times A_V
\label{regress}
\end{equation}
This result is compatible with a fractal structure of the cloud if 
$A_V \propto \log H$, i.e. if the density of matter follows a power law, which
is, precisely, what is used in modelling interstellar clouds \cite{BBP93}.

\subsection{Maximum of extinction}
\label{max_ext}
Fig. \ref{mass_spec} shows the relation between iso--extinction contours
and the logarithm of the mass contained inside these contours for the Taurus
cloud (see extinction map in Fig. \ref{taurus}). The relation is linear for
$A_V \la 5.5$. 
For higher extinction, mass is deficient because the star count leads to 
underestimating the extinction. Indeed, for highly extinguished regions, the
low number density of stars requires a larger area to pick up enough stars and 
estimate the extinction. The result is therefore an average value over a
large area in which the extinction is, in fact, greater. In the Taurus cloud,
I found that this turn off occurs for a size of $\sim 1.7$ pc. According to
Blitz and Williams \cite*{BW97} a turn off toward higher masses should appear
for a size scale of $\sim 0.5$ pc. Obviously, our value corresponds to a
limitation of the star counts method with optical data and not to a real 
characteristic size scale of the cloud.

It is therefore natural to extrapolate, down to a minimum mass, the linear part
of the relation $\log \left[ M(A_v) \right]$ versus $A_V$ to determine a 
maximum of extinction. This maximum is obtained using the regression line 
(\ref{regress}) and represents the higher extinction that can be measured,
would the cloud be fractal at all size scales. As Blitz and Williams 
\cite*{BW97} have shown, there is a characteristic size scale above which 
the density profile becomes steeper. The extrapolation of the linear relation
gives, therefore, a lower limit for the densest core extinction. Derived
values of $A_V$ are presented in the $4^{th}$ column of Table \ref{tab} and
correspond to a minimum mass of $1 M_{\sun}$. This minimum mass is a typical
stellar mass and, an extrapolation toward lower masses would be meaningless.

\noindent
Except for the Carina cloud, maxima are found in the range from 5.7
to 25.5 magnitudes of visual extinction with a median value of 10.6 magnitudes.

We stress the point that extinction can be larger. The $\rho$ Ophiuchus cloud
,for example, is known to show extinction peaks of about $\sim 100$ magnitudes 
\cite{CMFA95}, whereas we obtain only 25.5 magnitudes. Nevertheless, this value
compared to the 9.4 magnitudes effectively measured indicates that we need
deeper optical observations or near--infrared data --- such as those provided
by the DENIS survey --- to investigate more deeply the cloud. The Coalsack and
Scorpius are the only clouds for which measured and extrapolated extinctions
are similar (see Sect. \ref{remarks}).

\subsection{Mass}
Assuming a gas to dust ratio, the mass of a cloud can be obtained using
the relation \cite{Dic78}~:
\begin{displaymath}
M=(\alpha d)^2 \mu \frac{N_H}{A_V} \sum_i{A_V(i)}
\end{displaymath}
\noindent
where $\alpha$ is the angular size of a pixel map, $d$ the distance to the
cloud, $\mu$ the mean molecular weight corrected for helium abundance, and
$i$ is a pixel of the extinction map.
Uncertainties on the determination of masses come essentially from the
distance which is always difficult to evaluate. Assuming a correct distance
estimation, error resulting from magnitude uncertainties can be evaluated:
an underestimation of 0.5 magnitude of visual extinction implies a reduction
of the total mass of a factor $\sim 2$.
According to Savage and Mathis \cite*{SM79}, the gas to dust ratio is
$\frac{N_H}{A_V}~=~1.87 \times 10^{21} \, {\rm cm}^{-2}.{\rm mag}^{-1}$
where $N_H~=~N_{HI}+2N_{H_2}$. Kim and Martin \cite*{KM96} show that this
value depends on the total to selective extinction ratio $R_V = A_V/E_{B-V}$.
For $R_V = 5.3$, the gas to dust ratio would be divided by a factor 1.2.
The value of $R_V$ is supposed to be larger in molecular clouds than in the
general interstellar medium but the variation with the extinction is not
clearly established. So, I used the general value of 3.1 and the gas to dust
ratio of Savage and Mathis. Mass of the cores of the clouds may, therefore, be
overestimated by a factor $\sim 1.2$.

In Fig. \ref{mass_spec}, the relation $\log \left[ M(A_v) \right]$ vs. $A_V$
extrapolated toward the zero extinction gives an estimation of the total
mass of the cloud using Eq. (\ref{regress}). Masses obtained are shown in the
Table \ref{tab}. The median mass is $2900 M_{\sun}$ and the range is from 80
to $5 \, 10^5 M_{\sun}$. Using expression (\ref{regress}), we also remark that
half of the total mass is located outside the iso--extinction curve 1.0
magnitudes. This value is remarkably stable from cloud to cloud with a
standard deviation of 0.3. 

\begin{figure*}[htb]
\caption{Extinction map of Orion, Monoceros I, Rosette, Monoceros II, Crossbones from $R$ counts (J2000 coordinates)}
\label{orion}
\end{figure*}

\begin{figure*}[htb]
\caption{Extinction map of Taurus from $R$ counts (J2000 coordinates)}
\label{taurus}
\end{figure*}

\begin{figure*}[htb]
\caption{Extinction map of Serpens from $R$ counts (J2000 coordinates)}
\label{serpens}
\end{figure*}

\section{Remarks on individual clouds}
\label{remarks}
\subsection{Vela and Serpens}
In the Vela and the Serpens clouds (Fig. \ref{vela} and \ref{serpens},
respectively), there is no linear relation between 
$\log \left[ M(A_v) \right]$ and $A_V$. Extrapolation for the maximum of
extinction or for the total mass estimations is not possible. However, mass 
lower limits can be obtained using the extinction map directly:
$5.7 \, 10^4 \, M_{\sun}$ and $1.1 \, 10^5 \, M_{\sun}$ for Vela and
Serpens, respectively. It is difficult to understand why there is no linear 
relation for these two clouds, even for low values of extinction.

\subsection{Carina}
The study of the Carina (Fig. \ref{carina}) presents aberrant values for the
slope of the line $\log \left[ M(A_v) \right]$ (see Table \ref{tab}).
Consequently, the maximum of the extrapolated extinction, 82 magnitudes, cannot
be trusted. The important reflection in the Carina region is probably 
responsible for the shape of the extinction map. Stars cannot be detected 
because of the reflection and thus, extinction cannot be derived from $R$ star
counts. Infrared data are requested to eliminate the contribution of the
nebulae.

\subsection{Musca--Chamaeleon}
The Chamaeleon I has already been mapped with DENIS star counts in $J$ band
\cite{CEC+97} and the maximum of extinction was estimated to be $\sim 10$
magnitudes. Using $B$ star counts we find here 5.2 magnitudes. This difference 
is normal since $J$ is less sensitive to extinction than $B$. The important
remark is that the extrapolated value for the extinction derived from $B$ star
counts is 12.9, consistent with the value obtained with $J$ star counts.
We obtain the same result for the Chamaeleon II cloud for which $J$ star
counts lead also to a maximum of about $\sim 10$ magnitudes whereas the
extrapolated value from $B$ counts is 12.3 magnitudes. Infrared data are
definitely necessary to investigate the cores of the clouds.

Besides, the shape of the whole Musca--Chamaeleon extinction and the IRAS 
$100\mu m$ maps are very similar. 
Disregarding the far--infrared gradient produced by the heating by the 
galactic plane, there is a good match of the far--infrared emission and 
extinction contours and filamentary connections.
This region is well adapted to study the correlation
between the extinction and the far--infrared emission because there is no 
massive stars which heat the dust. The $100 \mu m$ flux can, therefore, be 
converted in a relatively straightforward way into a column density unit using
the $60/100 \mu m$ colour temperature \cite{BBDT98}.

\subsection{Coalsack and Scorpius}
The extinction map of the Coalsack is displayed in Fig. \ref{coalsack}. 
The edge of the cloud contains the brightest star of the Southern Cross 
($\alpha=12^h26^m36^s, \,\, \delta=-63\degr 05' 57''$).
This cloud is known to be a conglomerate of dust material, its distance
is therefore difficult to estimate.
Franco \cite*{Fra95} using Str\"omgren photometry gives a distance of
150-200 pc.
More recently Knude and Hog \cite*{KH98} estimate a distance of 100-150 pc
using Hipparcos data. 
Finally, I adopted an intermediate value of 150 pc to derive the
mass of the Coalsack.
Maximum extinction estimations for this cloud are 6.6 and 6.3 for the measured 
and the extrapolated values, respectively. 
The lowest limit for the maximum of extinction, as defined in Sect. 
\ref{max_ext}, is reached, but no characteristic size scale has been found by
studying the shape of $\log \left[ M(A_v) \right]$ at high extinction.
These values are too close, regarding their uncertainties, to
reflect any evidence of a clumpy structure.

Nyman et al. \cite*{NBT89} have made a CO survey of the Coalsack cloud. They
have divided the cloud into 4 regions. Regions I and II which correspond to the
northern part of the area near $\alpha$ Crux, are well
correlated with the extinction map. Nyman et al. have defined two other
regions
which have no obvious counterpart in extinction. Region III below 
-64\degr of declination in the western part of the cloud is not seen in the 
extinction map and this may result of a scanning defect of the plates.
The same problem
exists for the region IV which is a filament in the eastern part of the cloud
 at $\delta \simeq -64\degr$.

The Scorpius cloud is located near the $\rho$ Ophiuchus cloud (Fig.
\ref{rho_oph}). The 120 pc distance used to derive its mass is the $\rho$ 
Ophiuchus distance \cite{KH98}. As for the Coalsack cloud, measured and
extrapolated extinction are similar : 6.4 and 7.0, respectively. But, no 
evidence of a characteristic size scale can be found. 
For both clouds, this result is not surprising since the maximum of
measured extinction reaches the extrapolated value. Would dense cores with
steeper extinction profile be found, the measured extinction would have been
significantly greater than the extrapolated value.

\subsection{Corona Australis}
The extinction in this cloud has recently been derived using star counts on
$B$ plates by Andreazza and Vilas-Boas \cite*{AV96}. Our methods are very
similar and we obtain, therefore, comparable results. The main difference comes
from the most extinguished core. Since they use a regular grid, they cannot
investigate cores where the mean distance between two stars is greater than
their grid step. Consequently, they obtained a plateau where I find 4 distinct
cores.

\subsection{IC5146}
CO observations are presented in Lada et al. \cite*{LLCB94}. 
The two eastern cores in the $^{13}$CO map have only one counterpart in 
the extinction map because the bright nebula, Lynds 424, prevents the star
detections in that region.
Lada et al. \cite*{LLCB94} also present an extinction map of a part of the
cloud derived from $H-K$ colour excess observations. This colour is
particularly well adapted for such investigations because, in one hand,
infrared wavelengths allow deeper studies and, on the other hand, $H-K$ colour
has a small dispersion versus the spectral type of stars.
I obtain similar low iso--extinction contours but they reach a much greater 
maximum of visual extinction of about 20 magnitudes. 

\subsection{Lupus}
New estimations of distance using Hipparcos data \cite{KH98} have led to locate
the Lupus complex (Fig. \ref{lupus}) at only 100 pc from the Sun.
Lupus turns out to be the most nearby star--forming cloud.
This distance is used to estimate the complex mass but, it is important to
remark that evidence of reddening suggests that dust material is present 
up to a distance of about 170 pc \cite{Fra90}. Masses could therefore be
underestimated by a factor $\la 3$ in these regions.
The complex has been first separated into 4 clouds \cite{Sch77}, and then, 
a fifth cloud has been recently discovered using $^{13}$CO survey
\cite{TDM+96}. 
I have discovered, here, a sixth cloud which happens to be as massive as the
Lupus I cloud, $\sim 10^4 \, M_{\sun}$ (assuming it is also located at 100 pc).
The measured extinction for this cloud reaches 4.8 magnitudes.

The comparison between the $^{13}$CO and the extinction map is striking,
especially for the Lupus I cloud for which each core detected in the molecular
observations has a counterpart in extinction.
Mass estimations can be compared on condition that the same field is used
for both maps.
Moreover, $^{13}$CO observations are less sensitive than $B$ star counts for low
extinction. The lower contour in the $^{13}$CO map corresponds to a visual
extinction of $\sim 2$ magnitudes. Using this iso--extinction contour to
define the edge of the cloud and the same distance (150 pc) as Tachihara 
et al. \cite*{TDM+96}, I find a mass of $\sim 1300 M_{\sun}$ in agreement
with their estimation of $1200 M_{\sun}$.
Murphy et al. \cite*{MCM86} estimate the mass of the whole complex
to be $\sim 3 \, 10^4 M_{\sun}$ using $^{12}$CO observations and a distance
 of 130 pc. Using the same distance, I would obtain $4 \, 10^4 M_{\sun}$  
($2.3 \, 10^4 M_{\sun}$ for a 100 pc distance).

\subsection{$\rho$ Ophiuchus}
Because of the important star formation activity of the inner part of the
cloud, the IRAS flux at $100 \mu m$ shows different structures of those seen in
the extinction map. 
On the other hand, the 3 large filaments are present in both maps. Even if
these regions are complex because several stars heat them, the comparison 
between
the far--infrared emission and the extinction should allow to derive
the dust temperature and a 3-dimensional representation
of the cloud and of the stars involved in the heating. Unfortunately,
uncertainties about the distances for these stars are too large 
(about 15\%) and corresponds roughly to the cloud size.

\subsection{Orion}
Maddalena et al \cite*{MMMT86} have published a large scale CO map of Orion and
Monoceros R2. Masses derived from CO emission are consistent with those I
obtain : $1.9 \, 10^5 M_{\sun}$ and $0.86 \, 10^5 M_{\sun}$ for Orion and
Monoceros R2, respectively, from CO data and $3 \, 10^5 M_{\sun}$ and $1.2 \,
10^5 M_{\sun}$ from the extinction maps.
The Orion B maps looks very alike. The Orion A maps show a significant
difference near the Trapezium ($\alpha = 5^h35^m$, $\delta = -5\degr 23'$)
where the young stars pollute the star counts.
The correlation between CO and extinction maps for Monoceros R2 is less
striking, because of the star forming activity. 
It is clear that star clusters involve an underestimation of the extinction,
but I would like to stress the point that the heating by a star cluster may
also destroy the CO molecules. Estimation of the column density from star
counts and from CO observations may, therefore, be substantially underestimated
in regions such as the Trapezium. 

\subsection{Taurus}
Onishi et al. \cite*{OMK+96} have studied the cores in the Taurus
cloud using a C$^{18}$O survey. All of the 40 cores identified in their survey
are also detected in the extinction map. Moreover, Abergel et al.
\cite*{ABMF94} have shown a strong correlation between the far--infrared and
the $^{13}$CO emission in that region. Despite the complexity of
the Taurus structure (filaments, cores), it is a region, like the Chamaeleon
complex, located at high galactic latitude ($b \simeq -16\degr$), without
complex stellar radiation field. This situation is highly favourable to
a large scale comparison of CO, far--infrared and extinction maps. 

\section{Conclusion}
Star counts technique is used since the beginning of the century and is still a
very powerful way to investigate the distribution of solid matter in molecular
clouds. Now, with the development of digital data, this technique become easy 
to use and can probe much larger areas.
For all regions, we have assumed that all stars were background stars.
The error resulting from this hypothesis can easily be estimated. Equation 
(\ref{extinction}) can be written :
\begin{displaymath}
A_\lambda = \frac{1}{a} \log \left( \frac{S}{nb} \right) + cste(D_{ref})
\end{displaymath}

\noindent
where $S$ is the surface which contains $nb$ stars.
If 50\% of the stars are foreground stars, the difference between the real
extinction and the extinction which assumes all background stars is :
\begin{displaymath}
\Delta A_\lambda = \frac{1}{a} \log 2
\end{displaymath}

\noindent
That corresponds to $\sim 0.6$  or $\sim 1.1$ magnitudes of visual extinction 
whether star counts are done using $B$ or $R$ band, respectively.
Fortunately, most of the clouds are located at small or intermediate distances
to the Sun (except the Carina at 2500 pc) and this effect is probably
small, at least for low extinction. The good agreement between mass
derived from extinction or from CO data argues in that favour. 

\noindent
Stars physically associated to the clouds are more problematic because they are
located precisely close to the extinction cores. Young objects are generally
faint in the optical band so this problem may be neglected for optical counts.
This is no longer true with near--infrared data for which young objects must be
removed before the star counts.
We obtain extinction map with a spatial resolution always adapted to the local
densities which are typically about $\sim 1'$ for the outer part of cloud and
$\sim 10'$ for the most extinguished regions where the stellar density becomes
very low, i.e. $\la 1000 {\rm \ stars.deg^{-2}}$.
These maps allow the estimation of the total mass of the cloud by
extrapolation of the distribution of matter with the extinction. 
It appears that mass concentrated in the regions of low extinction represents
an important part of the total mass of a cloud (1/2 is contained in regions
of extinction lower than 1 magnitude).

Extrapolation of the distribution of matter in highly extinguished areas
is more risky. Star counts method give a relation for which masses are
underestimated in the cores of the cloud for a well understood reason:
estimation of the local density requires to pick up enough stars and thus, to
use larger area because of the low number density. 
Moreover a characteristic size scale in the distribution of matter \cite{BW97}
indicates the presence of the lower limit of the fractal cloud structure.
For this size scale the linear extrapolation used in Fig. \ref{mass_spec}
also underestimates the real mass and extinction. Despite this difficulty,
the extrapolated extinction is useful to estimate the {\em saturation} level
in the extinction map, but it is important to keep in mind that extinction can
be much larger in small cores. 

Finally, the examination of the relation between mass and extinction is useful
to check what we are measuring. The Carina cloud show an aberrant slope
(Table \ref{tab}) which is a strong indication that the map cannot be directly
interpreted as an {\em extinction} map. In that case, the elimination of
reflection nebulae is probably a solution and therefore, near-infrared data
are requested to investigate the extinction.
For the Vela and the Serpens cloud, the absence of linear part is not
understood and reflection is not the solution in these regions. 

\begin{acknowledgements}
I warmly thank N. Epchtein for initiating this study and for his critical
reading of the manuscript which helped to clarify this paper.
The {\em Centre de Donn\'ees astronomiques de Strasbourg} (CDS) is also
thanked for accessing to the USNO data.
\end{acknowledgements}

\end{document}